\documentstyle[epsfig,prc,aps,12pt]{revtex}

\begin {document}
\tighten
\title{Effect of continuum couplings in 
fusion of halo $^{11}$Be on $^{208}$Pb around the Coulomb barrier}
\author{A. Diaz-Torres$^{\thanks{%
E-mail: A.Diaz-Torres@surrey.ac.uk}}$ and I.J. Thompson}
\address{Physics Department, University of Surrey, Guildford GU2 7XH, 
United Kingdom}
\date{\today}
\maketitle

\begin{abstract}
The effect of continuum couplings in the fusion of the halo nucleus 
$^{11}$Be on $^{208}$Pb 
around the Coulomb barrier is studied using a 
three-body 
model within a coupled discretised continuum channels (CDCC) 
formalism. 
We investigate in particular the role of continuum-continuum couplings. 
These are found to hinder total, complete and incomplete fusion 
processes. Couplings to 
the projectile $1p_{1/2}$ bound excited state 
redistribute the complete and incomplete fusion 
cross sections, but the total fusion cross section remains nearly constant.
Results show that continuum-continuum couplings enhance the 
irreversibility of breakup and reduce the flux that penetrates 
the Coulomb barrier. 
Converged total fusion cross sections agree with the 
experimental ones for energies around the Coulomb barrier, but underestimate 
those for energies well above the Coulomb barrier. 
\end{abstract}



 
$\emph{Introduction:}$ The existence and the role of the breakup process of
weakly bound projectiles in complete fusion and scattering mechanisms have 
been extensively investigated in recent years
both theoretically \cite{TH2,New,TH1,Yabana,Dasso,hagino} and 
experimentally \cite
{Yoshida,Takahashi,Rehm,Signorini1,Kolata,Dasgupta,Trotta,Aguilera,Signorini2,Lubian}, but there is not yet any definitive conclusion. 
There are contradictory theoretical works which predict either the 
suppression \cite{TH2,New,TH1,Yabana} or the enhancement \cite{Dasso} of 
the complete fusion
cross section due to the coupling of the relative motion of the nuclei to 
the breakup channel.
 
Recent coupled
channels calculations for $^{11}$Be+$^{208}$Pb \cite{hagino} 
have shown that the coupling
of the relative motion to the breakup channel has two effects, depending on
the value of the bombarding energy, namely (i) a reduction of the complete
fusion cross sections at energies above the Coulomb barrier due to the loss
of incident flux, and (ii) an enhancement of the complete fusion cross sections
at energies below the Coulomb barrier due to the dynamical renormalisation
of the nucleus-nucleus potential. Using the isocentrifugal approximation and
an incoming boundary condition inside the barrier, this calculation did not
include the effect of the projectile's halo structure on 
the monopole projectile-target potential. Nor did it include the excitation 
to partial waves 
other than $p_{3/2}$ in the continuum, or the continuum-continuum and 
bound excited states couplings in either reaction partner. 
Moreover, only a small 
interval of energy for continuum states (up to 2 MeV) was considered. 

The couplings between continuum states have been shown to be crucial 
to understand the breakup of $^{8}$B on a $^{58}$Ni target at low 
energy $E_{lab}=25.8$ MeV \cite{CDCC3,CDCC4}. 
Therefore, it could be expected that continuum-continuum couplings 
significantly affect the role of breakup process in fusion of halo nuclei 
around 
the Coulomb barrier. We believe that continuum-continuum couplings enhance 
the irreversibility of the breakup process (thus once the projectile gets 
dissociated, it will find it 
very hard to find its way back to the bound states). 
Thus, the continuum-continuum couplings are expected to reduce the flux in 
bound projectile channels and, therefore, should inhibit (at least) 
the complete fusion. We expect that, with an increasing breakup subspace, 
governed by both the maximum energy and continuum partial waves, 
continuum-continuum couplings will reduce the fusion cross sections 
until convergence is reached.

The aim of this paper is to clarify the role of these continuum couplings on 
the fusion of the halo
projectile $^{11}$Be on a $^{{208}}$Pb target at energies around 
the Coulomb barrier. 
Calculations are carried out using a three-body model \cite{CDCC3,CDCC4} in 
the framework of the CDCC formalism \cite{CDCC1,CDCC2}. Full coupled channels 
calculations are performed with the code FRESCO \cite{Fresco}. 

In calculation of fusion cross sections, we simultaneously include 
(i) the effect of the projectile's halo structure on 
the projectile-target potential, (ii) both the transition to its bound
excited state and its dissociation caused by inelastic excitations to 
different partial waves in the
continuum, induced by the projectile fragments-target interactions (coulomb
+ nuclear), and (iii) couplings (bound-continuum and continuum-continuum) 
between its excited states. We do not include 
transfer or inelastic channels of the
target. Fusion cross sections for projectile bound channels 
and for projectile breakup channels will be 
defined in terms of a short-ranged imaginary bare potential defined in the
center of mass coordinate of the projectile in conjunction with channel 
dependent wave functions for the projectile-target radial motion. 
 


 
$\emph{Model:}$ In the $^{11}$Be + $^{208}$Pb reaction, the
three-bodies involved are the $^{10}$Be core ($C$), the valence 
(halo) neutron ($v$), and the $^{208}$Pb target ($T$).

Let $\overrightarrow{R}$ be the coordinate from the target to the center of
mass of the projectile, and $\overrightarrow{r}$ the internal coordinate of
the projectile. The position coordinates of the projectile fragments with
respect to the target are 
$\overrightarrow{r}_{vT}=\overrightarrow{R}+\frac{A_{P}-1}{A_{P}}%
\overrightarrow{r}$
and 
$\overrightarrow{R}_{CT}=\overrightarrow{R}-\frac{1}{A_{P}}%
\overrightarrow{r}$, 
where $A_{P}$ is the mass of the projectile.

The dynamics of the three-bodies is described by the Schr\"{o}dinger equation
in the over-all centre of mass system. 
The two-body potentials $%
V_{CT}$, $V_{vT}$ and $V_{vC}$ depend only on relative coordinates indicated
as their arguments and do not excite the internal degrees of freedom of the
core and the target nucleus. Following \cite{hagino}, these potentials are 
considered as real, but in addition we use for fusion 
a bare short-ranged (well inside the Coulomb barrier) imaginary central
potential $iW_{F}(R)$ defined in the center of mass of the projectile for
the projectile-target radial motion. 
The use of this short-ranged imaginary potential is equivalent
to the use of an incoming boundary condition inside the Coulomb barrier to 
study fusion \cite{Brown,Landowne}.

In order to describe the breakup of a projectile such a $^{11}$Be, we 
consider the inelastic excitations induced by $V_{CT}$, $V_{vT}$ in the n + $%
^{10}$Be system from the ground state $\phi _{(ls)j,n}^{g.s}(r)$ to excited
states in the continuum $u_{(ls)j,k}(r)$, for some wave-number $k$ and partial
wave $l$, and also couplings between all such continuum states. The use of 
such single energy eigenstates, however, would result
in calculations of form factors for continuum-continuum couplings which 
do not converge, as
the continuum wave functions are not square integrable. 
The CDCC method \cite{CDCC1,CDCC2} 
is used to obtain square integrable continuum 
bins states $\phi _{(ls)j,[k_{1},k_{2}]}(r)$ averaged over a narrow range of
wave-numbers $[k_{1},k_{2}]$. We label these bin states by their wave-number
limits $[k_{1},k_{2}]$ and their angular momentum quantum numbers $(ls)j$.
The bound states of the projectile $\phi _{(ls)j,n}(r)$ and the single
energy scattering wave functions $u_{(ls)j,k}(r)$ which form the continuum
bins $\phi _{(ls)j,[k_{1},k_{2}]}(r)$, are obtained 
by solving a 
Schr\"{o}dinger equation with the potential $V_{vC}^{l}$ which may be $l$%
-dependent. The bin wave functions are defined as
\begin{equation}
\phi _{(ls)j,[k_{1},k_{2}]}(r)=\sqrt{\frac{2}{\pi N}}\int\limits_{k_1}^{k_2}
w(k)e^{-i\delta_{k}}u_{(ls)j,k}(r)dk,  
\label{bin}
\end{equation}
where $\delta_{k}$ is the scattering phase shift for $u_{(ls)j,k}(r)$. 
The normalisation constant is 
$N=\int\limits_{k_1}^{k_2}\vert w(k)\vert^{2}dk$ for the assumed 
weight function $w(k)$, here taken to be either unity for non-$s$-wave bins or 
$k$ for $s$-wave bins. 
These bin states are normalised $\langle\phi\vert\phi\rangle =1$ once a 
sufficiently large maximum radius 
$r_{bin}$ for $r$ is taken. They are orthogonal to any bound states, and 
are orthogonal to other bin states if their energy ranges do not overlap. 
The phase factor $e^{-i\delta_{k}}$ ensures that they are real valued 
for real potentials $V_{vC}^{l}$.     

The radial wave functions $f_{\alpha J}(R)$ for the projectile-target 
relative motion satisfy the set of coupled equations \cite{CDCC3} 
\[
\left[ -\frac{\hbar ^{2}}{2\mu }\left( \frac{d^{2}}{dR^{2}}-\frac{L(L+1)}{%
R^{2}}\right) +V_{\alpha :\alpha }^{J}(R)+iW_{F}(R)+\epsilon _{\alpha
}-E\right] f_{\alpha J}(R)
\]
\begin{equation}
=\sum_{\alpha ^{\prime }\neq \alpha }i^{L^{\prime }-L}V_{\alpha :\alpha
^{\prime }}^{J}(R)f_{\alpha ^{\prime }J}(R),  \label{NEq}
\end{equation}
where $\mu$, $L$, $J$, $E$ and $\alpha $ ($\{L,l,s,j,n$ or $[k_{1},k_{2}]\}$)
denote the projectile-target reduced mass, the projectile orbital angular 
momentum, the total angular momentum,
the total energy, and the set of quantum numbers, respectively. 
For unbound states of the projectile, $\epsilon _{\alpha }$ is
the mean energy of continuum bin $[k_{1},k_{2}]$ 
, or $\epsilon _{\alpha
}<0$ for bound states. $V_{\alpha :\alpha ^{\prime }}^{J}$ describes the
coupling between the different internal states $\phi _{\alpha }(%
\overrightarrow{r})$ of the projectile 
\begin{equation}
V_{\alpha :\alpha ^{\prime }}^{J}(\overrightarrow{R})=<\phi _{\alpha }(\overrightarrow{r}%
)\mid V_{CT}(\overrightarrow{R}_{CT})+V_{vT}(\overrightarrow{r}_{vT}
)\mid \phi _{\alpha ^{\prime }}(\overrightarrow{r})>.  \label{Eq6}
\end{equation}

Assuming that the potentials $V_{CT}$ and $V_{vT}$ are central, the Legendre
multipole potentials can be formed as 
\begin{equation}
\Lambda _{K}(R,r)=\frac{1}{2}\int\limits_{-1}^{+1}[V_{CT}(\overrightarrow{R}%
_{CT})+V_{vT}(\overrightarrow{r}_{vT})]P_{K}(x)dx,  \label{Eq7}
\end{equation}
where $K$ is the multipole and $x=\widehat{r}\cdot \widehat{R}$ is the
cosine of the angle between $\overrightarrow{r}$ and $\overrightarrow{R}$.
Since the spin $s$ of the neutron is fixed, the coupling form factor (\ref
{Eq6}) between states $\phi _{\alpha ^{\prime }}(r)$ and $\phi _{\alpha }(r)$
is 
\begin{eqnarray}
V_{\alpha :\alpha ^{\prime }}^{J}(R) &=&\sum_{K}(-1)^{j+j^{\prime }-J-s}%
\widehat{j}\widehat{j}^{\prime }\widehat{l}\widehat{l}^{\prime }\widehat{L}%
\widehat{L}^{\prime }(2K+1)W(jj^{\prime }ll^{\prime };Ks)W(jj^{\prime
}LL^{\prime };KJ)  \nonumber \\
&&\times \left( 
\begin{array}{lll}
K & l & l^{\prime } \\ 
0 & 0 & 0
\end{array}
\right) \left( 
\begin{array}{lll}
K & L & L^{\prime } \\ 
0 & 0 & 0
\end{array}
\right) \int\limits_{0}^{\infty }\phi _{\alpha }(r)\Lambda _{K}(R,r)\phi
_{\alpha ^{\prime }}(r)dr.  \label{Eq9}
\end{eqnarray}

Eqs. (\ref{NEq}) are solved with the usual scattering boundary conditions 
\cite{Fresco}.


The total fusion cross section $\sigma _{tot}$ is defined in terms of that
amount of flux which leaves the coupled channels set because of the
short-ranged imaginary potential $iW_{F}(R)$.

Since complete fusion is a process where all the nucleons of the projectile
are captured by the target nucleus and following \cite{hagino}, 
we define in our model the complete fusion cross
section as the absorption cross section from projectile bound channels
(complete fusion from both ground state (elastic) and bound excited state) 
\begin{equation}
\sigma _{CF}=\frac{\pi }{2\mu \text{E}}\sum_{J}(2J+1)P_{J},  \label{Eq10}
\end{equation}
where E is the bombarding energy and $P_{J}$ is the complete fusion 
probability for the partial wave $J$. The complete fusion probability $P_{J}$ 
is \cite{Satchler} 
\begin{equation}
P_{J}=\frac{8}{\hbar (2\text{E}/\mu )^{1/2}}\sum_{\alpha \text{ }(\epsilon
_{\alpha }<\text{ }0)}\int\limits_{0}^{\infty }\mid f_{\alpha J}(R)\mid
^{2}\lgroup-W_{F}(R)\rgroup dR.  \label{Eq11}
\end{equation}

The complete fusion cross section (\ref{Eq10})-(\ref{Eq11}) represents a 
lower limit for the physical complete fusion cross section, 
since we have assumed no capture of all projectile fragments 
($^{10}$Be and the halo neutron) from breakup channels. 
In reality, these events should contribute to the complete fusion, but 
cannot be distinguished in our model from the capture of only one projectile 
fragment.

The incomplete fusion $\sigma _{ICF}$ (fusion of $^{10}$Be) is then 
defined as the absorption from breakup channels 
     


$\emph{Results and discussion:}$ The experimental spectrum of $^{11}$Be 
exhibits 
a 1/2$^{+}$ ground state and a single, 
1/2$^{-}$, bound excited state with energies of --0.50 MeV and --0.18 MeV, 
respectively. In a pure single-particle picture, the ground and 
the bound excited states of $^{11}$Be have $2s_{1/2}$ and $1p_{1/2}$ 
single-particle configurations, respectively. These configurations 
can be associated with single-particle states generated by different 
$V_{vC}^{l}$ Woods-Saxon potentials \cite {Lenzi} including 
a spin-orbit term. 
For the $2s_{1/2}$ state, we use a Woods-Saxon potential with parameters 
$V_{0}$ = --51.51 MeV, $r_{0}$ = 1.39 fm and $a$ = 0.52 fm. 
For the $1p_{1/2}$ state, we use a 
Woods-Saxon potential including a spin-orbit term, similar to that used in 
\cite{Lenzi}, with the same geometry, 
i.e. $V_{0}$ = --30 MeV, $r_{0}$ = 1.39 fm, $a$ = 0.52 fm 
and $V_{0}^{s.o.}$ = 4.39 MeV.      

First, we study qualitatively the effect of continuum couplings on fusion 
cross sections by using a reduced breakup subspace with regard to the maximum 
energy of the projectile continuum states. A continuum breakup subspace 
with partial waves 
$s_{1/2}$, $p_{1/2}$, $p_{3/2}$, $d_{3/2}$ and $d_{5/2}$, for 
the halo neutron-$^{10}$Be core relative motion, is used. 
For each partial wave, the continuum 
subspace is discretised in 6 bins which are equally spaced in wave-number $k$, 
up to a maximum wave-number $k_{max}=0.3612$ fm$^{-1}$ 
(a maximum energy of 3 MeV), 
with a step of $\Delta k=0.0602$ fm$^{-1}$. In Fig.1, we illustrate 
the continuum discretisation used to define the energy bins 
included in these calculations.  
The calculation is thus performed with 30 excited 
continuum channels. The $s$- and $p$-wave continuum states have been 
consistently generated by the same potential $V_{vC}^{l}$ 
as that of the bound state of the same angular momentum $l$. The $d$-wave 
continuum states have been generated by the same potential as that of the 
$p$-waves.

In the present work, Woods-Saxon parametrisations given in \cite{Broglia1} and 
in \cite{Mottelson} are used for the nuclear part of 
the potentials $V_{CT}$ ($V_{0}$ = --46.764 MeV, $r_{0}$ = 1.192 fm 
and $a$ = 0.63 fm) 
and $V_{vT}$ ($V_{0}$ = --44.019 MeV, $r_{0}$ = 1.27 fm and $a$ = 0.67 fm), 
respectively. A short-ranged 
Woods-Saxon potential $W_{F}$ with 
parameters $V_{0}$ = --50 MeV, $r_{0}$ = 1 fm and $a$ = 0.1 fm is used for 
the fusion potential. 
The results depend only weakly on the geometry of this potential, 
as long as it is well inside 
the Coulomb barrier and strong enough that the mean-free path of 
the projectile inside the barrier is much smaller than 
the dimensions of $W_{F}$. The fusion cross sections 
for $V_{0}$ = --50 MeV are those for $V_{0}$ = --10 MeV changed 
by $\sim$ 1\%.
 
Since we are interested in fusion cross sections, partial waves for 
the projectile-target relative motion up to only $L_{max}$ = 50 
(partial-wave total fusion cross section $\sim$ 10$^{-3}$ mb) are included. 
Our calculations 
include monopole, dipole and quadrupole contributions ($K$=0,1 and 2) of 
the potentials 
$V_{CT}$ and $V_{vT}$ for both nuclear and Coulomb parts. The couplings 
$V_{\alpha:\alpha ^{\prime}}^{J}(R)$ are taken into account up to a 
projectile-target radial distance $R_{coup}$ = 100 fm. 
To calculate both 
the continuum bins (\ref{bin}) and couplings 
$V_{\alpha:\alpha ^{\prime}}^{J}(R)$ (\ref{Eq9}), which include these 
bins, radii $r\leq r_{bin}=100$ fm are used. 
 
  
Figs.2a and 2b show fusion cross sections as a function of the 
bombarding energy in the center of mass system. For comparison, 
we present cross sections in 
the absence of couplings (thin solid curve). 
In Fig.2a, calculations include transitions from and to the 
projectile bound states, 
but do not include 
continuum-continuum couplings. In this case, we found that 
the effect of 
the projectile bound excited state $1p_{1/2}$ on the total, complete and 
incomplete fusion cross sections is 
quite weak ($\sim$ 10\%). The couplings to the bound excited state 
$1p_{1/2}$ only redistributes the complete fusion cross section 
(thick solid curve) 
between the elastic channel and this channel, the fusion contribution from the 
elastic channel being $1.7-3.4$ times larger than from the bound excited state 
$1p_{1/2}$ for the range of energies studied. We would like 
to note that these 
fusion excitation functions show similar trends as those obtained by 
Hagino et al. \cite{hagino}. We agree that complete fusion cross sections 
are strongly enhanced due to the couplings to the projectile excited states 
compared with the no-coupling case at energies below and just above the 
Coulomb barrier ($V_{B} \approx$ 36 MeV for the elastic channel), whereas 
they are hindered at above barrier energies. 

In Fig.2b, we show the effect of continuum-continuum couplings 
on the total and complete fusion cross sections of Fig. 2a. 
It is found that 
well above the Coulomb barrier, both total and complete fusion 
cross sections are suppressed 
compared with the no-coupling case, and enhanced well below the barrier. 
Just below the Coulomb barrier 
(34 MeV $\leq$ E$_{\text{c.m.}}$ $\leq$ 36 MeV), complete fusion 
cross sections 
are suppressed, but this is not the case for total fusion cross sections. 
In the present case, we found that couplings to the projectile bound 
excited state 
$1p_{1/2}$ redistribute (dot-dashed curve) 
the complete and incomplete fusion cross sections, 
while the total fusion cross sections (dashed curve) remain nearly 
constant. 
With couplings to the bound excited state $1p_{1/2}$, 
the contribution to complete fusion 
from the elastic channel is similar to the one from the bound excited state 
$1p_{1/2}$ for energies below the Coulomb barrier, and 
$1.7-8$ times smaller for energies above the Coulomb barrier.  

Fig.3 shows incomplete fusion excitation functions 
(difference between the total and the complete fusion curves) for both 
cases presented above, namely in Figs.2a and 2b. We can observe that 
continuum-continuum couplings significantly reduce 
the incomplete fusion cross sections (dashed curve).  
The case when the couplings to the bound excited 
state $1p_{1/2}$ are not included is shown by the dot-dashed curve.

From Fig.2b and Fig.3, it is observed that continuum-continuum couplings 
strongly affect the predicted total, complete and incomplete 
fusion cross sections. This implies that the fusion dynamics strongly 
depends on continuum-continuum couplings. Since the short-ranged 
imaginary potential is well confined within the Coulomb barrier, we deduce 
that continuum-continuum couplings mainly reduce the flux that penetrates the 
barrier, while couplings to the projectile bound excited state $1p_{1/2}$ 
mainly redistribute, among the complete and incomplete 
fusion channels, the flux that has already penetrated the Coulomb barrier.   

We have checked the convergence of reported fusion cross sections 
(total, complete and incomplete) 
with the size of the breakup subspace, and have found the following when 
couplings between all projectile excited states (bound-continuum and 
continuum-continuum) are included in the calculation:
\begin{itemize}
\item  The maximum energy of the continuum states (Fig.4): 
a maximum energy beyond 9 MeV is needed to obtain converged results. 
With respect to the fusion cross sections for a maximum energy of 9 MeV 
(dashed curve), 
fusion cross sections for a maximum energy of 10 MeV (full squares) 
are changed by 
$\sim$ 10\% for energies 
around the Coulomb barrier. For energies well above the barrier, fusion cross 
sections are changed by $\sim$ 1.5\%.
 
\item  The density of the continuum discretisation (Fig.5):
a density greater than 1.67 bins/MeV is needed to obtain converged results. 
The same density is used for all partial waves in the continuum. 
With respect to fusion cross sections for a density of 1.67 bins/MeV 
(dotted curve), 
fusion cross sections for a density of 2 bins/MeV (dashed curve) 
are changed by 
$\sim$ 6.5\% for energies 
around the Coulomb barrier. For energies well above the barrier, fusion cross 
sections are changed by $\sim$ 1.6\%.

\item  The number of partial waves in the continuum 
and potential multipoles (Fig.6): 
partial waves beyond $f_{5/2}$, $f_{7/2}$ and 
potential multipoles beyond the octupole contribution ($K$=3) are needed to 
obtain converged results. 
With respect to fusion cross sections for continuum 
partial waves up to $f$-waves and potential 
multipoles $K \leq$ 3 (full circles), 
fusion cross sections for continuum partial waves up to $g$-waves 
and potential multipoles $K \leq$ 4 (full triangles) 
are changed by 
$\sim$ 8\% (total fusion), $\sim$ 100\% (complete fusion) and 
$\sim$ 3\% (incomplete fusion), respectively, for energies 
around the Coulomb barrier. For energies well above the barrier, fusion cross 
sections are changed by 
$\sim$ 13\%. The calculation including both continuum partial waves up to 
$g$-waves and potential 
multipoles $K \leq$ 4 (full triangles) is presently at 
the limit of our computational capability.
\end{itemize}
 
Fig.7 shows experimental total fusion cross sections (full squares) for 
the similar 
system $^{11}$Be + $^{209}$Bi \cite{Signorini1}, which should not differ 
too much from the reaction studied. By comparing converged total 
fusion cross sections for $^{11}$Be + $^{208}$Pb (full stars), calculated 
within our model, with the experimental ones for $^{11}$Be + $^{209}$Bi, 
it is observed that the converged total fusion excitation function does not 
reproduce the experimental one. They do agree 
with the experiment for energies around the Coulomb barrier, but 
underestimate the data by $\sim$ 41\% for energies well above the Coulomb 
barrier. 

A crude estimation of the effect of target excitations on the total 
fusion cross section has been done by (i) fitting 
the converged total fusion cross section in a single (elastic) channel 
calculation by finding an appropriate projectile-target real Wood-Saxon 
potential with an energy dependent depth and 
the geometry $r_{0}$ = 1.179 fm and $a$ = 0.658 fm, and then (ii) including 
the target excitations as in 
ref. \cite{thomp2}. Such estimation reveals that the effect is quite weak. 
Fusion cross sections are increased $\sim$ 1.28 times for energies around 
the Coulomb barrier, while they remain nearly constant for energies 
well above the Coulomb barrier.
 
The experimental cross sections for  $^{11}$Be + $^{209}$Bi were 
obtained \cite{Signorini1} as the sum of three channels: 5n+4n+fission. 
It was pointed out in ref. \cite{newsign} that the 3n channel, expected to be 
relevant below the barrier, could not be measured and at the same time 
the fission cross section could have been overestimated. 
A new experiment is necessary in order to clarify 
the $^{11}$Be fusion mechanism discussed in the present work.        


$\emph{Summary and conclusions:}$ Fusion cross sections 
calculated in the CDCC framework depend strongly on continuum-continuum 
couplings.
We do not include transfer or inelastic channels of the target. 
Continuum-continuum couplings 
hinder total, complete and incomplete fusion processes. 
Couplings to the projectile $1p_{1/2}$ 
bound excited state redistribute the 
complete and incomplete fusion cross sections, but do not change 
the total fusion cross section. Results show that continuum-continuum 
couplings enhance the irreversibility of breakup and reduce the flux 
that penetrates the Coulomb barrier. 
A large breakup subspace is needed to 
obtain converged fusion cross sections. The converged total fusion excitation 
function does not reproduce the experimental one: 
converged total fusion cross sections agree with the 
experimental ones for energies around the Coulomb barrier, but underestimate 
those for energies well above the Coulomb barrier. A crude estimation of 
the effect of target excitations on the total fusion cross section reveals 
that it is quite weak. A new experiment seems 
to be necessary to clarify the $^{11}$Be fusion mechanism discussed 
in the present work. 
The total fusion cross section is unambigously calculated in our formalism, 
but this is not the case for the complete fusion since the capture of 
all projectile fragments from breakup channels cannot always be 
distinguished from the capture of only one projectile fragment.
 
{\bf Acknowledgments}

We thank Prof. Jeff Tostevin for a careful reading of the paper, 
helpful discussions 
and comments. We also thank Profs. Cosimo Signorini and Andrea Vitturi 
for fruitful discussions. We are grateful to Dr. Atsushi Yoshida for 
the experimental data. 
UK support from the EPSRC grant GR/M/82141 is acknowledged.


\newpage

\begin{figure}
\begin{center}
\epsfig{file=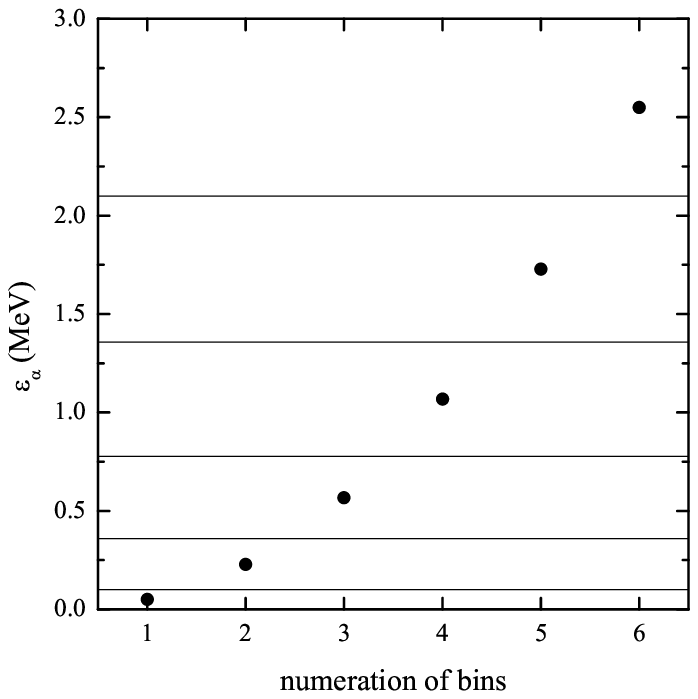,width=0.75\textwidth}
\end{center}
\caption{Continuum discretisation used to define the energy bins. 
The central energies of bins are shown by full circles.}
\end{figure}

\newpage

\begin{figure}
\begin{center}
\epsfig{file=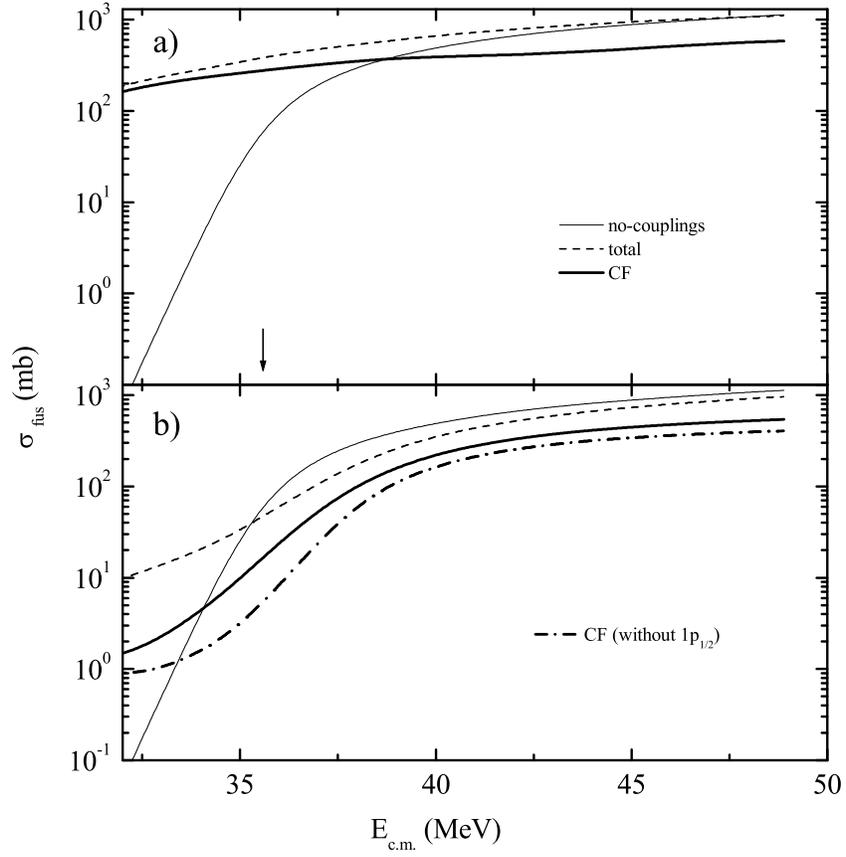,width=0.75\textwidth}
\end{center}
\caption{Fusion cross sections as a function of the bombarding energy in 
the center of mass system for $^{11}$Be + $^{208}$Pb. a) Include only 
couplings from and to the $^{11}$Be bound states. 
b) Couplings between all $^{11}$Be excited states (continuum-continuum) 
are included. See text for further details. 
The arrow indicates the Coulomb barrier for the elastic channel. }
\end{figure}

\newpage

\begin{figure}
\begin{center}
\epsfig{file=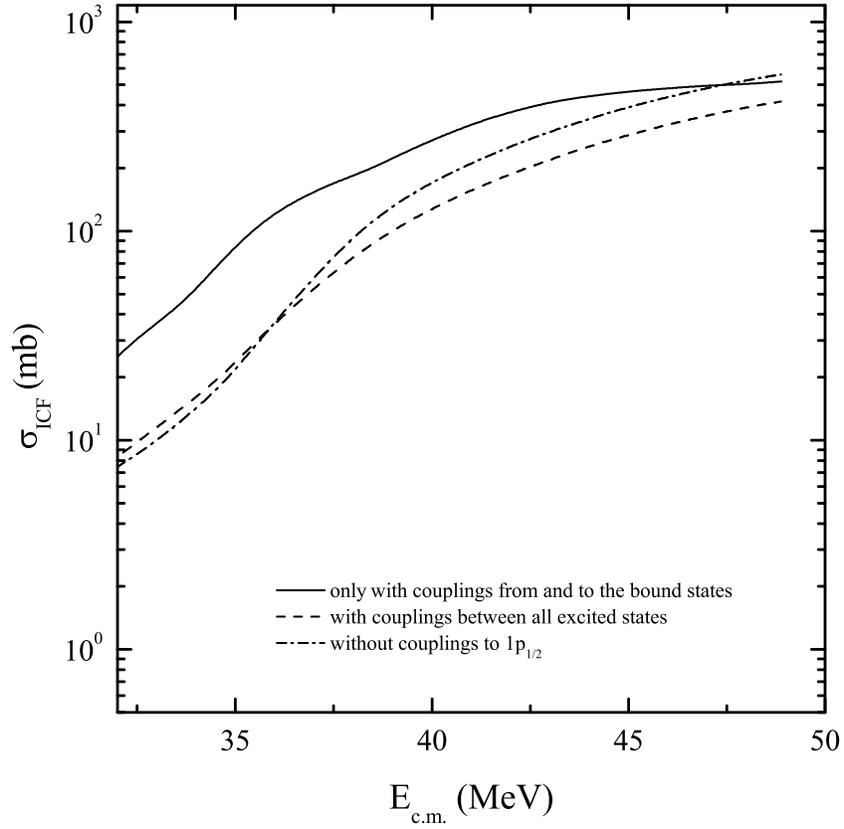,width=0.75\textwidth}
\end{center}
\caption{Incomplete fusion excitation functions for both cases shown in 
Figs.2a (solid curve) and 2b (dashed curve). Incomplete fusion excitation 
function for the case of Fig.2b, but couplings to the $^{11}$Be bound 
excited state $1p_{1/2}$ are not included (dot-dashed curve). 
}
\end{figure}

\newpage

\begin{figure}
\begin{center}
\epsfig{file=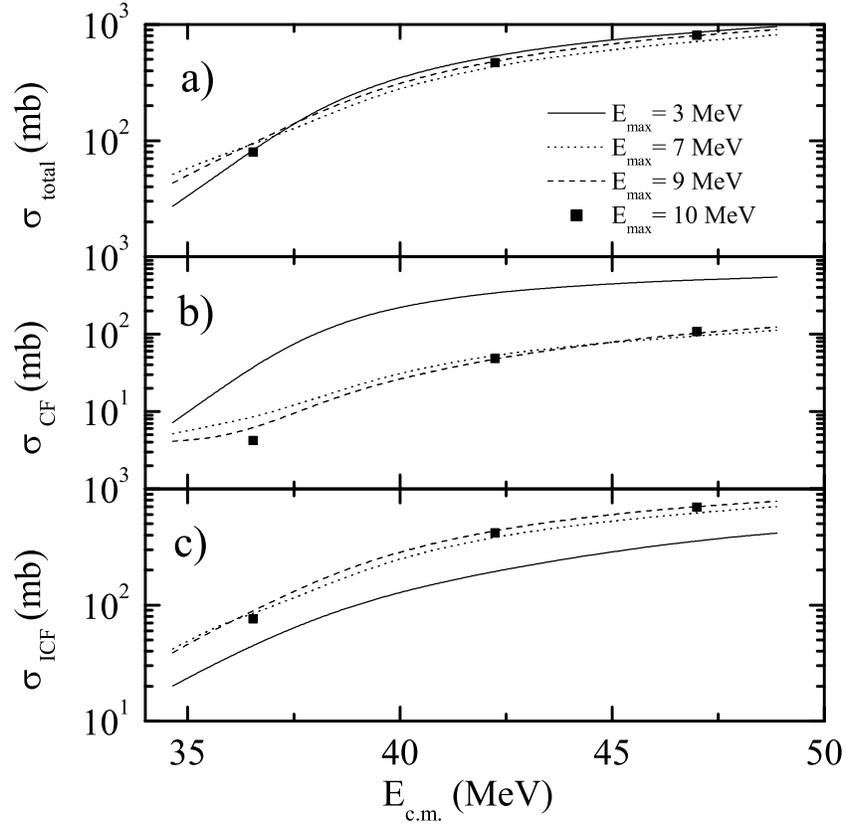,width=0.75\textwidth}
\end{center}
\caption{Convergence of total (a), complete (b) and incomplete (c) fusion 
excitation functions for $^{11}$Be + $^{208}$Pb with regard to the maximum 
energy of the $^{11}$Be continuum states included in the calculation. 
The $s$-, $p$- and $d$-wave continuum states for a density of the continuum 
discretisation of 2 bins/MeV, potential multipoles $K \leq$ 2 and 
couplings between $^{11}$Be excited states are 
included in the calculation. See text for further details.
}
\end{figure}

\newpage

\begin{figure}
\begin{center}
\epsfig{file=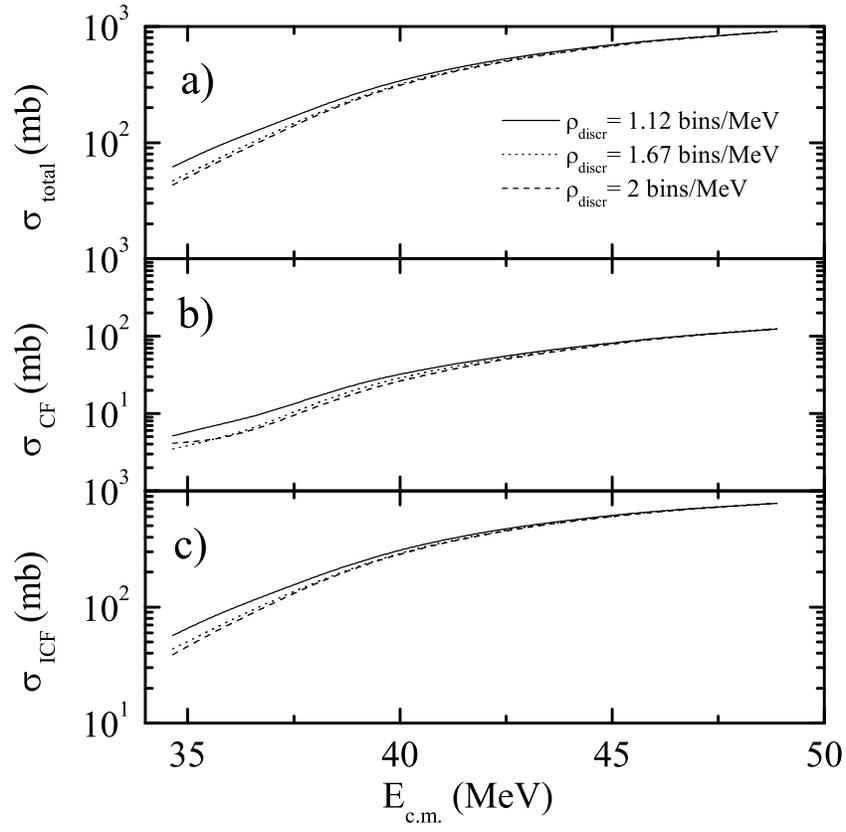,width=0.75\textwidth}
\end{center}
\caption{The same as in Fig.4, but with regard to the density of the continuum 
discretisation. The maximum energy of the $^{11}$Be continuum states is 9 MeV.}
\end{figure}

\newpage

\begin{figure}
\begin{center}
\epsfig{file=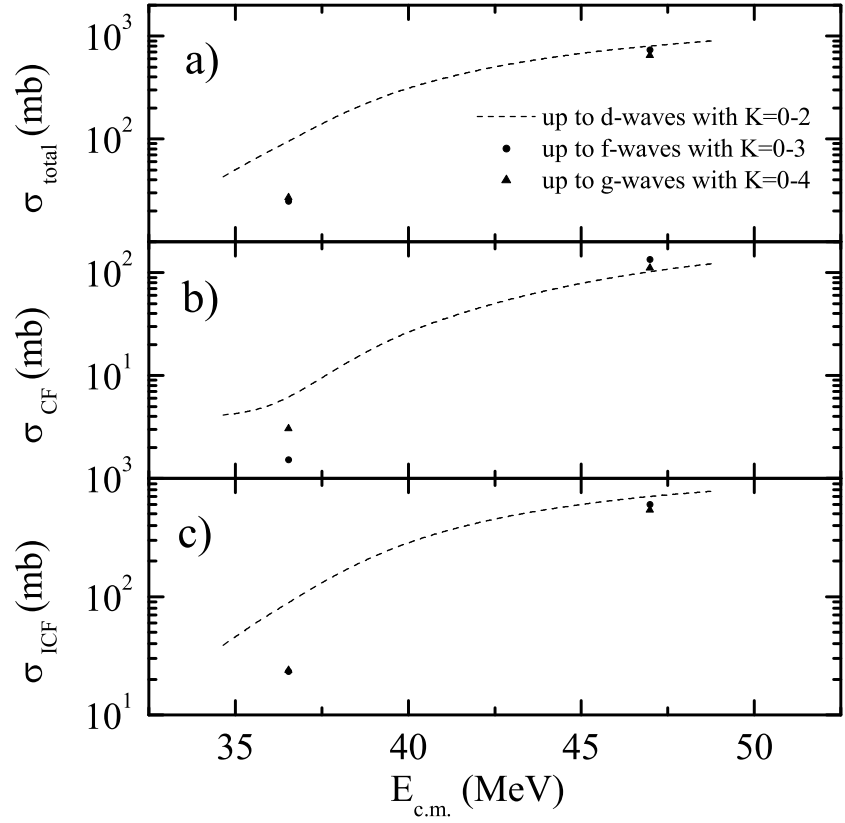,width=0.75\textwidth}
\end{center}
\caption{The same as in Figs.4 and 5, but with regard to the number of partial 
waves in the continuum and potential multipoles.
The maximum energy of the $^{11}$Be continuum states and 
the density of the continuum discretisation are 9 MeV and 2 bins/MeV, 
respectively.}
\end{figure}

\newpage

\begin{figure}
\begin{center}
\epsfig{file=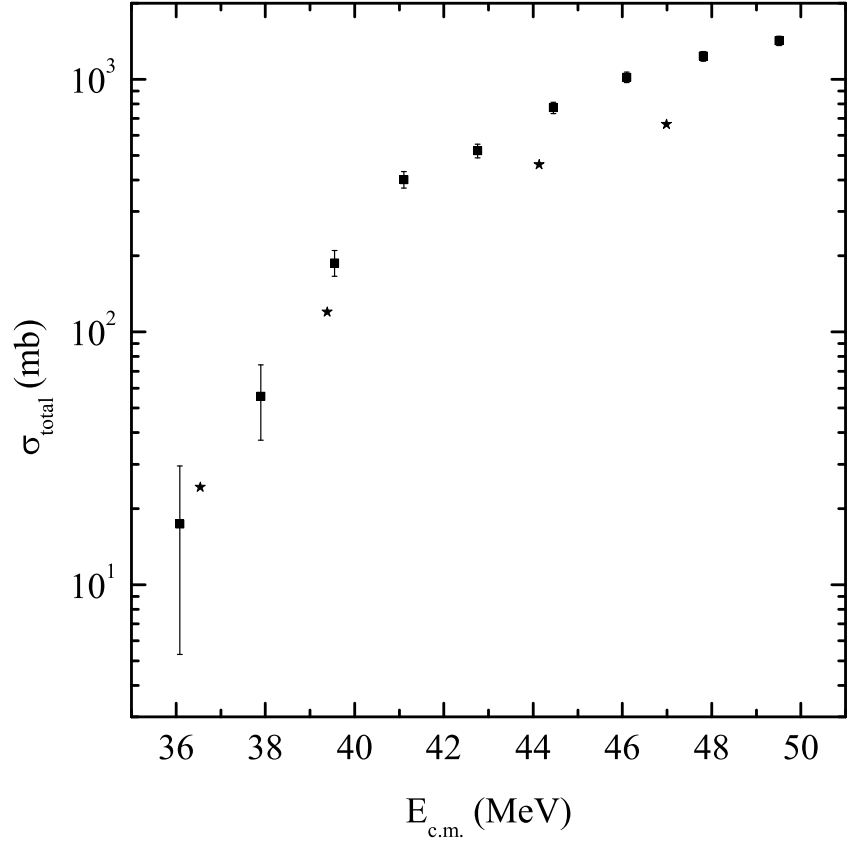,width=0.75\textwidth}
\end{center}
\caption{Converged total fusion cross sections 
for $^{11}$Be + $^{208}$Pb (full stars) are compared with 
the experimental ones \protect\cite{Signorini1} 
for $^{11}$Be + $^{209}$Bi (full squares). 
A maximum energy of the $^{11}$Be continuum states of 10 MeV, 
continuum partial waves up to $g$-waves for a density of the continuum 
discretisation of 2 bins/MeV, potential multipoles $K \leq$ 4 and 
couplings between $^{11}$Be excited states are included in the calculation.}
\end{figure}


\end{document}